\documentstyle[12pt]{article}

\begin{document}

\begin{titlepage}

\begin{flushright}
ENSLAPP xxx/93\\
\end{flushright}
\par \vskip .1in \noindent

\begin{center}
{\LARGE  On Pentagon And Tetrahedron Equations}\\
 \end{center}
  \par \vskip .3in \noindent

\begin{center}

      {\bf Jean Michel MAILLET }
  \par \vskip .1in \noindent

{\sl  Laboratoire de Physique Th\'eorique $^{*}$\\
       ENS Lyon, 46 all\'ee d'Italie 69364 Lyon CEDEX 07
       France}\\[0.8in]
\end{center}

\par \vskip .40in
\begin{center}
{\bf Abstract}\\
\end{center}

\begin{quote}
We show that solutions of Pentagon equations lead to solutions of the
Tetrahedron equation. The result is obtained in the spectral parameter
dependent case.
\end{quote}
\par \vskip 1.8in

\begin{flushleft}
\rule{5.1 in}{.007 in}\\
$^{*}${\small URA 1436 ENSLAPP du CNRS, associ\'ee \`a  l'Ecole
Normale Sup\'erieure de Lyon et au Laboratoire d'Annecy de Physique
des Particules. }\\
{\small { email: maillet@enslapp.ens-lyon.fr}}\\[0.2 in]

Ref. ENSLAPP xxx/93 \\
November 1993
\end{flushleft}

\end{titlepage}

{\bf 1. Introduction}\\
\\

Yang-Baxter (or triangle, or $2$-simplex) equations \cite{Yang,Bax1}, play a
central role in the theory of two-dimensional Integrable Systems of Field
Theory and Statistical Mechanics (see for reviews \cite{Bax1,F}). They also
lead to the theory of Quantum Groups \cite{Drin1,Jim1,Jim2,Skl1,KR1,FRT} and
have
important applications in low dimensional Topology. In 1980,
A. B. Zamolodchikov \cite{Zam1,Zam2} described a generalization of this
equation, the Tetrahedron (or $3$-simplex) equation, for three-dimensional
Integrable Systems. This equation can be further extended to an arbitrary
dimension $d$ and is called the $d$-simplex equation \cite{BS1}. More recently,
the first solution of the Tetrahedron equation, proposed in \cite{Zam2} (see
\cite{Bax2} for the proof), has been generalized using results from the
two-dimensional Chiral Potts models \cite{BB1,KMS}.\\

The purpose of this letter is to give a construction of unitary solutions
of the Tetrahedron equation (depending on spectral parameters) in terms
of  solutions of  Pentagon equations. Our starting point is the
geometrical interpretation of these
equations given in \cite{JMM1}. It is argued in \cite{JMM1} that the
$d$-simplex equation can be obtained as a special discretized
case of a (generalized) zero holonomy equation for transport operators acting
in a space of functionals of $(d-1)$-dimensional manifolds. In this picture,
an $R$-matrix $R_{d}$ solving the $d$-simplex equation is associated to a
$d$-dimensional parallelepipedic cell, and is  interpreted as an operator
moving a functional of $d$ of its $2d$ faces to a functional of the $d$ other
faces. The condition for parallel transport (zero holonomy) is then
precisely the $d$-simplex equation. For $d=1$ it gives Lax type equations,
for $d=2$ the Quantum Yang-Baxter equation, for $d=3$ the Tetrahedron
equation, etc. \\

In short, this equation can be described as follows. Let $\Sigma_{d}$ and
$\Sigma_{d}^{'}$ be two oriented  $d$-dimensional manifolds having the
same compact oriented boundary which is divided into two $(d-1)$-dimensional
oriented manifolds $\Sigma_{d-1}^{*}$ ($\Sigma^{*}$ meaning the same manifold
as $\Sigma$ but with reversed orientation) and $\Sigma_{d-1}^{'}$ having
also the same boundary. Then we associate to $\Sigma_{d-1}$ and
$\Sigma_{d-1}^{'}$ respectively two vector spaces $V_{\Sigma_{d-1}}$
and $V_{\Sigma_{d-1}^{'}}$. Let $F (\Sigma_{d})$ be a map,
\begin{equation}
 F (\Sigma_{d})\ :\  V_{\Sigma_{d-1}}\  \longmapsto\ V_{\Sigma_{d-1}^{'}}
\label{eq:F}
\end{equation}
Then  $F (\Sigma_{d})$ can be interpreted as a transport operator
(depending on the manifold  $\Sigma_{d}$) acting on  functionals
of $(d-1)$-dimensional manifolds . The condition for parallel transport is
just,
\begin{equation}
F({\Sigma}_{d})\ =\ F({\Sigma}_{d}^{'})
\label{eq:FFn}
\end{equation}
for any two manifolds $\Sigma_{d}$ and $\Sigma_{d}^{'}$ satisfying the above
conditions, in particular, $\partial \Sigma_{d}\ =\ \partial \Sigma_{d}^{'}$.\\

However as noticed
in \cite{JMM1}, it is also possible to give another discrete version of
such a (generalized) zero holonomy equation in terms of operators ${\Phi}_{d}$
attached to $d$-simplices instead of $d$-cells. Equations of this type for
${\Phi}_{d}$'s are called the Fundamental $(d+1)$-Simplex Relations (due to
the fact that they are written around a $(d+1)$-simplex, each face of it being
a $d$-simplex associated to one ${\Phi}_{d}$ and that they realize
eq. (\ref{eq:FFn}) in the minimal (simplicial) way.). Moreover, the operator
$R_{d}$ attached to any $d$-cell can be obtained as an (ordered) product of
the $d!$ operators ${\Phi}_{d}$ attached to  $d!$ $d$-simplices in which this
$d$-cell can be decomposed. Given such a formula, the $d$-simplex equation
for $R_{d}$ is a consequence of the fundamental $(d+1)$-simplex relations for
the ${\Phi}_{d}$'s.\\

For $d=2$ this procedure gives the decomposition of a quantum
$R$-matrix in terms of $F$ type objects satisfying quadratic equations (a
$3$-simplex possesses four faces, one $F$ being attached to each of them).
In that case, from the algebraic
point of view, this procedure gives the geometrical interpretation of
the construction, used by V. G. Drinfel'd in \cite{Drin2}, of solutions of
the Quantum Yang-Baxter equation.\\
In \cite{Drin2}, unitary solutions $R_{12}$, namely
$R_{12}(u,v)\ R_{21}(v,u)\ = {\bf 1}$,  of the Quantum Yang-Baxter equation,
\begin{eqnarray}
&R_{12}(u,v)\  R_{13}(u,w)\
R_{23}(v,w)\ =\nonumber\\
&=\ R_{23}(v,w)\ R_{13}(u,w) \ R_{12}(u,v)
\label{eq:yb}
\end{eqnarray}
are obtained in terms of a more fundamental object $F_{12}$ such that,
\begin{equation}
R_{12}(u,v)\ =\ F_{21}^{-1}(v,u)\ F_{12}(u,v)
\label{eq:RF}
\end{equation}
($(u,v)$ being two vectors spectral parameters), $R_{12} \in A \otimes A$, $A$
being a Hopf algebra with co-commutative co-product ${\Delta}_{0}$,
$F_{12} \in A \otimes A$ and the $F$ objects satisfy the quadratic
($3$-simplex) relation,
\begin{equation}
({\Delta}_{0} \otimes {\bf 1})F\ F_{12}\ =\ ({\bf 1} \otimes {\Delta}_{0})F
\ F_{23}
\label{eq:FF2}
\end{equation}\\

The geometrical setting for this construction and its generalizations to
non-unitary cases is given in \cite{JMM2}. It is used in \cite{FM1} to
construct from any given classical $r$-matrix $r \in {\cal G} \otimes {\cal G}$
 the corresponding universal quantum $R$-matrix as a functional of $r$,
together with the quantized Hopf (quasi-triangular) algebra $\cal A$,
$R \in {\cal A} \otimes {\cal A}$.\\

Then, for $d=3$, the $R$-matrix $R(u,v,w)$ is interpreted as an
operator associated to
a three-dimensional parallelepipedic cell (depending on three vectors
$(u,v,w)$), and acting in a space of functionals of surfaces. The condition
for parallel transport is the Tetrahedron equation.
A three-dimensional parallelepipedic
cell can be decomposed into six tetrahedrons (and one "hat", see below). We
associate one $\Phi$ to each tetrahedron (and one $\Gamma$ to the "hat"), such
that the $R$-matrix decomposes as a product of six $\Phi$'s and one $\Gamma$.
Then the $4$-simplex relation for $\Phi$ is in fact a Pentagon equation.
The zero holonomy requirement imposes also some consistency relations
between $\Phi$ and $\Gamma$. These relations for $\Phi$ and $\Gamma$ imply
that the $R$-matrix $R(u,v,w)$ satisfies the Tetrahedron equation. This will
be our main result.\\
All equations will be given here for vertex models,
namely for indices attached to surfaces (plaquettes or triangles). A
completely similar description exists for variables on links or on points
(or for all these possibilities together) and will be described elsewhere
as well as more details on proofs and examples.\\

This letter is organized as follows. In section 2, we define the objects $R$,
$\Phi$, and $\Gamma$ in the three-dimensional case and give their geometrical
meaning together with the decomposition of $R$ in terms of $\Phi$ and $\Gamma$.
 We also give  the Tetrahedron equation for $R$. In section 3, we descibe
Pentagon equations in this geometrical framework. Our main result is stated
in section 4. There we give the skeleton of the proof of the relation
between solutions of Pentagon and Tetrahedron equations. Perspectives and
conclusions are given in section 5.\\
\\
It is a great pleasure to dedicate this paper to L. D. Faddeev on the occasion
of his $60^{th}$ birthday.\\
\\
\\

{\bf 2. The Tetrahedron equation}\\
\\

For vertex models, the Tetrahedron equation can be written as follows
\cite{Zam1,Zam2,Bax2,JMM1,MN1},
\begin{eqnarray}
R_{123}(u,v,w)\ R_{145}(u,v,t)\ R_{246}(u,w,t)\ R_{356}(v,w,t)\ =\nonumber \\
=\ R_{356}(v,w,t)\ R_{246}(u,w,t)\ R_{145}(u,v,t)\ R_{123}(u,v,w)
\label{eq:t}
\end{eqnarray}
where, $R_{ijk} \in End( V_{i} \otimes V_{j} \otimes V_{k} )$, $V_{i}$ being
vector spaces of dimensions $N_{i}$ and $u,v,w,t$ are four arbitrary vectors
(say elements of ${\bf C}^{n}$) parametrizing the $R$-matrices.\\

As sketched in the Introduction, such $R$-matrices can be interpreted as
transport operators on a space of functionals of surfaces. So let us first
describe this functional space in a discretized case. \\
We consider an $n$-dimensional affine space on ${\bf C}$, with origin $O$.
We denote by ${\Delta}^{(x)}(u,v)$ (or equivalently
${\Delta}^{(x+u)}(v,-u-v)$ or ${\Delta}^{(x+u+v)}(-u-v,u)$),
$(u,v,x)$ being vectors in ${\bf C}^{n}$,
the oriented triangle defined by the point $(O+x)$ and its oriented boundary
$(u,v,-u-v)$. To such a triangle we associate a vector (which is a
functional of this triangle) $h^{(x)}(u,v) \in V_{u}^{(x)}
\otimes V_{v}^{(x+u)} \otimes V_{-u-v}^{(x+u+v)} \otimes
{\cal A}_{(u,v)}^{(x)}$, where $V_{u}^{(x)}$ is a vector space attached to
the oriented link starting at point $(O+x)$ in direction $u$, such that its
dual vector space $V_{u}^{(x)*}$ is equal to $V_{-u}^{(x+u)}$, and
${\cal A}_{(u,v)}^{(x)}$ is a vector space attached to the triangle
${\Delta}^{x}(u,v)$. Here also, the dual vector space to
${\cal A}_{(u,v)}^{(x)}$ is ${\cal A}_{(u+v,-v)}^{(x)}$ associated to the
same triangle but with reversed orientation. Note also that we have,
${\cal A}_{(u,v)}^{(x)} \equiv {\cal A}_{(v,-u-v)}^{(x+u)} \equiv
{\cal A}_{(-u-v,u)}^{(x+u+v)}$.\\
We define a composition law for two $h$-functionals whenever the two
corresponding triangles have (at least) one edge in common with opposite
orientation by the evaluation of one $h$ on the other using the duality
bracket on the vector spaces attached to the common edge which are dual to
one another. For example, to any two-dimensional parallelepiped
${\Box}^{(x)}(u,v)$ starting at point $(O+x)$ with oriented boundary
$(u,v,-u,-v)$ we associate a functional
\begin{equation}
l^{(x)}(u,v)\ =\ {< h^{(x)}(u,v) , h^{(x)}(u+v,-u) > }_{V^{(x)}_{u+v}}
\label{eq:l}
\end{equation}
where we have used the natural duality bracket between $V^{(x)}_{u+v}$ and
its dual vector space denoted by ${< . , . > }_{V^{(x)}_{u+v}}$.
There, $l^{(x)}(u,v)$ is an element of the tensor product,
$V_{u}^{(x)} \otimes V_{v}^{(x+u)} \otimes V_{-u}^{(x+u+v)} \otimes
V_{-v}^{(x+v)} \otimes {\cal A}_{[u,v]}$, where ${\cal A}_{[u,v]}^{(x)}$
stands for
${\cal A}_{(u,v)}^{(x)} \otimes {\cal A}_{(u+v,-u)}^{(x)}$ and we will
require for simplicity ${\cal A}_{[u,v]}^{(x)}$ not to depend on the
vector $x$. Then it is also
possible to define the composition law for $l$-functional using their
decomposition in terms of the $h$'s.\\
As a useful example we consider the functionals
$$
j^{(x)}(u,v,w)\ =\ < l^{(x)}(v,u) , l^{(x+v)}(w,u) , l^{(x)}(w,v)
 >_{V_{u}^{(x+v)} \otimes V_{v}^{(x)} \otimes V_{w}^{(x+v)}}
$$
and
$$
k^{(x)}(u,v,w)\ =\ < l^{(x+u)}(w,v) , l^{(x)}(w,u) , l^{(x+w)}(v,u)
>_{V_{u}^{(x+w)} \otimes V_{v}^{(x+u+w)} \otimes V_{w}^{(x+u)}}
$$
Then, we define the operator $R^{(x)}(u,v,w) \in  End(\ {\cal A}_{[v,u]}
\otimes {\cal A}_{[w,u]} \otimes {\cal A}_{[w,v]}\ )$ as the map,
\begin{equation}
R^{(x)}(u,v,w) : j^{(x)}(u,v,w) \longmapsto k^{(x)}(u,v,w)
\label{eq:R}
\end{equation}
Here $R^{(x)}(u,v,w)$ is a functional of the parallelepipedic
three-dimensional cell at point $(O+x)$ defined by the three vectors
$(u,v,w)$.\\
We further impose a unitarity condition on this operator, namely, that the
map  $R^{(x+u+v+w)}(-u,-v,-w)$ is the inverse map to $R^{(x)}(u,v,w)$. Then,
by considering the two (minimal) ways of mapping the functional,
\begin{eqnarray}
< l^{(x)}(v,u) , l^{(x+v)}(w,u) , l^{(x)}(w,v) ,\nonumber\\
l^{(x+w+v)}(t,u) ,l^{(x+w)}(t,v) , l^{(x)}(t,w) >
\label{eq:s1}
\end{eqnarray}
where the duality bracket evaluation is on,
\begin{eqnarray}
V_{u}^{(x+v)} \otimes V_{v}^{(x)} \otimes V_{w}^{(x+v)} \otimes
V_{t}^{(x+v+w)}\nonumber\\
\otimes V_{t}^{(x+w)} \otimes V_{w}^{(x)} \otimes V_{v}^{(x+w)}
\otimes V_{u}^{(x+v+w)}
\label{eq:v1}
\end{eqnarray}
to the functional,
\begin{eqnarray}
< l^{(x+w+t)}(v,u) , l^{(x+t)}(w,u) , l^{(x+u+t)}(w,v) ,\nonumber\\
 l^{(x)}(t,u) , l^{(x+u)}(t,v) , l^{(x+u+v)}(t,w) >
\label{eq:s2}
\end{eqnarray}
where the duality bracket evaluation is on,
\begin{eqnarray}
V_{u}^{(x+w+t)} \otimes V_{v}^{(x+u+w+t)} \otimes V_{w}^{(x+u+t)}
\otimes V_{t}^{(x+u)}\nonumber\\
\otimes V_{t}^{(x+u+v)} \otimes V_{w}^{(x+u+v+t)} \otimes V_{v}^{(x+u+t)}
\otimes V_{u}^{(x+t)}
\label{eq:v2}
\end{eqnarray}
we obtain the following parallel transport condition on the $R$-matrices,
\begin{eqnarray}
R^{(x+t)}(u,v,w)\ R^{(x)}(u,v,t)\ R^{(x+v)}(u,w,t)\ R^{(x)}(v,w,t)\
=\nonumber\\
=\ R^{(x+u)}(v,w,t)\ R^{(x)}(u,w,t)\ R^{(x+w)}(u,v,t)\ R^{(x)}(u,v,w)
\label{eq:tf}
\end{eqnarray}
If we consider the simplified case where the operator $R^{(x)}(u,v,w)$ do not
depend on the shift $(x)$, we obtain the Tetrahedron equation (\ref{eq:t}),
the convention being that the vector spaces ${\cal A}_{[v,u]}$ are label
by numbers ${1,2,3,4,5,6}$ or better here ${(11'),(22'),...}$ with the
correspondence, $[v,u] \equiv (11')$ ($(v,u) \equiv (1)$ and
$(u,v) \equiv (1')$) and so on , $[w,u] \equiv 22'$, $[w,v] \equiv 33'$,
$[t,u] \equiv 44'$, $[t,v] \equiv 55'$ and $[t,w] \equiv 66'$. Then we have,
$$
R_{11',22',33'}^{(x)}(u,v,w)\ =\ R^{(x)}(u,v,w)
$$
The (local) unitarity condition is now,
\begin{equation}
R_{11',22',33'}^{(x)}(u,v,w)\ R_{1'1, 2'2, 3'3}^{(x+u+v+w)}(-u,-v,-w)\ =\
{\bf 1}
\label{eq:UR}
\end{equation}
Note here the exchange of spaces $(i)$ and $(i')$. The (local) Tetrahedron
equation is given by,
\begin{eqnarray}
R_{11',22',33'}^{(x+t)}(u,v,w)\ R_{11',44',55'}^{(x)}(u,v,t)\
R_{22',44',66'}^{(x+v)}(u,w,t) \ R_{33',55',66'}^{(x)}(v,w,t)\ =\nonumber \\
=\ R_{33',55',66'}^{(x+u)}(v,w,t)\ R_{22',44',66'}^{(x)}(u,w,t)\
R_{11',44',55'}^{(x+w)}(u,v,t) \ R_{11',22',33'}^{(x)}(u,v,w)\nonumber\\
\label{eq:T}
\end{eqnarray}
for any set of vectors $(u,v,w,t,x)$.\\

Let us now define two other transport operators $\Phi$ and $\Gamma$ as the
mappings,
\begin{eqnarray}
{\Phi}^{(x)}(u,v,w)\  :\  < h^{(x)}(u,v) , h^{(x)}(u+v,w)
 >_{V^{(x)}_{u+v}}\nonumber\\
\longmapsto < h^{(x)}(u,v+w) , h^{(x+u)}(v,w) >_{V^{(x+u)}_{v+w}}
\label{eq:fi}
\end{eqnarray}
and,
\begin{eqnarray}
{\Gamma}^{(x)}(u,v,w)\ :\ < h^{(x)}(u,v) , h^{(x)}(u+v,-v)
 >_{V^{(x)}_{u} \otimes V^{(x+u)}_{v}}\nonumber\\
\longmapsto < h^{(x)}(u+v+w,-w) , h^{(x)}(u+v,w)
 >_{V^{(x)}_{u+v+w} \otimes V^{(x+u+v)}_{-w}}
\label{eq:gamma}
\end{eqnarray}
where, ${\Phi}^{(x)}(u,v,w)$ is a linear map from
${\cal A}^{(x)}_{(u,v)} \otimes {\cal A}^{(x)}_{(u+v,w)}$ to
${\cal A}^{(x)}_{(u,v+w)} \otimes {\cal A}^{(x+u)}_{(v,w)}$. Similarly,
${\Gamma}^{(x)}(u,v,w)$ is a map from
${\cal A}^{(x)}_{(u,v)} \otimes {\cal A}^{(x+u+v)}_{(-v,-u)}$ to
${\cal A}^{(x)}_{(u+v+w,-w)} \otimes {\cal A}^{(x)}_{(u+v,w)}$. Moreover, for
simplicity, we will make the identifications (in the above formula for
$\Phi$ and $\Gamma$),
${\cal A}^{(x)}_{(u,v)} \equiv {\cal A}^{(x)}_{(u,v+w)}$,
${\cal A}^{(x)}_{(u+v,w)} \equiv {\cal A}^{(x+u)}_{(v,w)}$ for $\Phi$ and
similarly for $\Gamma$,
${\cal A}^{(x)}_{(u,v)} \equiv {\cal A}^{(x)}_{(u+v,w)}$ and
${\cal A}^{(x+u+v)}_{(-v,-u)} \equiv {\cal A}^{(x)}_{(u+v+w,-w)}$ such that
$\Gamma$ contains a permutation operator in its definition.\\

Using these operators it is quite easy to decompose the action of the
$R$-matrix in terms of $\Phi$ and $\Gamma$.\\
For this purpose, we put indices on $\Phi$ and $\Gamma$  to make
explicit the vector spaces they are acting upon, namely, using the
above conventions and identifications of vector spaces, we obtain for
example,
${\Phi}^{(x+v+w)}(u,-u-w,-v) \equiv {\Phi}_{23'}^{(x+v+w)}(u,-u-w,-v)$,
and so on.\\
We have,
\begin{eqnarray}
R_{11',22',33'}^{(x)}(u,v,w)\ =\ P_{12}\ P_{13'}\ P_{2'3'}\ P_{13}\
{\Phi}_{31}^{(x)}(w,u+v,-v)\nonumber\\
{\Phi}_{3'1'}^{(x)}(u+w,v,-v-w)\ {\Phi}_{32}^{(x)}(w,v,u)\ P_{1'2'}\
{\Phi}_{2'1'}^{(x)}(u+v+w,-w,-v)\nonumber\\
{\Gamma}_{13'}^{(x)}(v,u+w,-v)\ P_{12'}\ {\Phi}_{12'}^{(x+u+v)}(-u-v,v,u+w)\
{\Phi}_{23'}^{(x+v+w)}(u,-u-w,-v)\nonumber\\
\label{eq:RFG}
\end{eqnarray}
Note that in this formula each $\Phi$ is associated to one of the six
tetrahedrons decomposing the three-dimensional cell corresponding to the
$R$-matrix and having always the two points $(O+x)$ and $(O+x+u+v+w)$ among
their four vertices. These six tetrahedrons are labelled by the six possible
ordered triplets $(a,b,c)$, $a,b,c \in \{u,v,w\}$. Note also that the role of
$\Gamma$ is to create the only vertex of the three-dimensional cell
$(O+x, u,v,w)$, namely the point $(O+x+u+w)$, not present in the initial
surface.\\
\\

{\bf 3. The Pentagon Equation}\\
\\

We are now interested in writing the general equation (\ref{eq:F}) for the
operators $\Phi$ and $\Gamma$.\\

Let us first note the useful symmetry relations,
\begin{equation}
{\Phi}^{(x)}_{ij}(u,v,w)\ =\ {\Phi}^{(x+u+v)}_{ji}(w,-u-v-w,u)
\label{eq:sfi}
\end{equation}
and for $\Gamma$,
\begin{equation}
{\Gamma}^{(x)}_{ij}(u,v,w)\ =\ {\Gamma}^{(x+u+v)}_{ji}(-v,-u,u+v+w)
\label{eq:sg}
\end{equation}
Then we impose the unitarity relation on  $\Phi$,
\begin{equation}
{\Phi}^{(x+u)}_{ij}(v,w,-u-v-w)\ P_{ij}\ {\Phi}^{(x)}_{ij}(u,v,w)\ =\ {\bf 1}
\label{eq:ufi}
\end{equation}
and on $\Gamma$,
\begin{equation}
{\Gamma}^{(x)}_{ij}(u,v,w)\ {\Gamma}^{(x)}_{ji}(u+v+w,-w,-v)\ =\ {\bf 1}
\label{eq:ug}
\end{equation}
We also ask for the following composition law,
\begin{equation}
{\Gamma}^{(x+u+v)}_{ij}(w,-u-v-w,t)\ {\Gamma}^{(x)}_{ij}(u,v,w)\ =\
P_{ij}\ {\Gamma}^{(x)}_{ij}(u,v,t-u-v)
\label{eq:cg}
\end{equation}
It means in particular that ${\Gamma}^{(x)}_{ij}(u,v,-v)\ =\ P_{ij}$.\\

To obtain the $4$-simplex fundamental relation on $\Phi$, we consider the
two minimal ways of mapping the functional,
$$
< h^{(x)}(u,v) , h^{(x)}(u+v,w) , h^{(x)}(u+v+w,t)
 >_{V^{(x)}_{u+v} \otimes V^{(x)}_{u+v+w}}
$$
to the functional,
$$
< h^{(x)}(u,v+w+t) , h^{(x+u)}(v,w+t) , h^{(x+u+v)}(w,t)
 >_{V^{(x+u)}_{v+w+t} \otimes V^{(x+u+v)}_{w+t}}
$$
This gives the following Pentagon equation on $\Phi$,
\begin{eqnarray}
{\Phi}^{(x)}_{12}(u,v,w+t)\ {\Phi}^{(x)}_{23}(u+v,w,t)\ =\nonumber\\
{\Phi}^{(x+u)}_{23}(v,w,t)\ {\Phi}^{(x)}_{13}(u,v+w,t)\
{\Phi}^{(x)}_{12}(u,v,w)
\label{eq:P}
\end{eqnarray}

Then using again discretized versions of eq. (\ref{eq:F}), we obtain the
following constraints between $\Phi$ and $\Gamma$,
\begin{eqnarray}
{\Phi}^{(x+u+v+w)}_{12}(t,-u-v-w-t,u+v)\ {\Gamma}^{(x)}_{13}(u,v+w,t)
\nonumber\\
{\Phi}^{(x+u+v)}_{23}(w,-v-w,-u)\ =\nonumber\\
=\ {\Phi}^{(x+u+v)}_{23}(w+t,-t,-u-v-w)\ {\Gamma}^{(x+u+v)}_{12}(-v,v+w,t)
\nonumber\\
{\Phi}^{(x+u)}_{13}(v+w,-u-v-w,u+v)\ =\nonumber\\
=\ {\Phi}^{(x+u+v)}_{13}(w,-u-v-w,u+v+w+t)\ {\Gamma}^{(x)}_{23}(u,v,w+t)
\nonumber\\
{\Phi}^{(x+u+v+w)}_{12}(-u-v-w,u,v)\nonumber\\
\label{eq:fg}
\end{eqnarray}
and similarly,
\begin{eqnarray}
P_{14}\ P_{23}\ {\Gamma}^{(x)}_{12}(u,v,w)\  {\Gamma}^{(x)}_{34}(u+v+w,-w,-v)
\ =\  {\Phi}^{(x+u)}_{23}(-u,u+v+w,-w)\nonumber\\
{\Phi}^{(x+u)}_{24}(-u,u+v,w)\ {\Phi}^{(x)}_{13}(u,v+w,-w)\
{\Phi}^{(x)}_{14}(u,v,w)\nonumber\\
\label{eq:ffg}
\end{eqnarray}
This last relation is quite interesting since it allows us to compute the
operator $\Gamma$ in terms of $\Phi$'s up to its trace.\\
\\

{\bf 4. Tetrahedron equation from Pentagon Equation}\\
\\

We can now state our main result using all the above ingredients,\\

{\bf Theorem :}\\

{\em Let ${\Phi}_{ij}(u,v,w)$ and ${\Gamma}_{ij}(u,v,w)$ be defined by eqs.
(\ref{eq:fi},\ref{eq:gamma}) and satisfying eqs. (\ref{eq:sfi},\ref{eq:sg},
\ref{eq:ufi},
\ref{eq:ug},\ref{eq:cg},\ref{eq:P},\ref{eq:fg},\ref{eq:ffg}).
Then the $R$-matrix $R_{11',22',33'}^{(x)}(u,v,w)$ defined in terms of
$\Phi$ and $\Gamma$ in eq. (\ref{eq:RFG}) satisfies the Tetrahedron
equation (\ref{eq:T}) and is unitary.}\\

The proof of this theorem is quite long and will not be given here. In
particular it uses $24$-times the Pentagon equation for $\Phi$. So, instead
of giving an explicit proof, let us describe the main idea leading to its
construction. In fact this idea holds for any of the $d$-simplex equation
with regards to their relation to the corresponding fundamental
$(d+1)$-simplex relations when writing the $R$-matrix $R_{d}$ in terms of the
simplicial objects ${\Phi}_{d}$. Hence for simplicity let us explain it first
in the two-dimensional case, namely for the Yang-Baxter equation.\\

Geometrically, the $(l.h.s)$ of the quantum Yang-Baxter equation
(\ref{eq:yb}) is
associated to  three faces of a cube, namely to the surface corresponding
to the functional $k^{(x)}(u,v,w)$, while the $(r.h.s)$ is associated to the
three other faces of the cube, hence to the surface corresponding to the
functional $j^{(x)}(u,v,w)$. So the $(l.h.s)$ and the $(r.h.s)$ of the Quantum
Yang-Baxter equation can be viewed as related by the (symbolic) action of
$R(u,v,w)$.Then in a similar way, the operator $\Phi$ can be consider in the
two-dimensional case as the symbolic action  relating the $(l.h.s)$ and the
$(r.h.s)$ of eq. (\ref{eq:FF2}) for $F$, $\Gamma$ being related to unitarity
relation for $F$. Now the proof of the Quantum Yang-Baxter equation
(\ref{eq:yb}) from eqs. (\ref{eq:RF},\ref{eq:FF2}) is precisely given by the
decomposition of $R(u,v,w)$ in terms of $\Phi$ and $\Gamma$. Namely, to each
$\Phi$ corresponds the use of eq. (\ref{eq:FF2}) for two definite $F$'s, and
to $\Gamma$ corresponds the use of the unitarity for $F$. Indead, using
eq. (\ref{eq:FF2}) six-times, in the precise order given by the
(non-abelian) decomposition (\ref{eq:RFG}), and once the unitarity for $F$, we
can achieve the proof of eq. (\ref{eq:yb}).\\

Generalizing this procedure to the $d=3$ case amounts to decompose an
$R$-matrix attached to a four-dimensional cell into its $24$ $4$-simplices.
This gives the precise way to use $24$-times the Pentagon equation to prove
the Tetrahedron equation for $R(u,v,w)$. In fact to achieve the proof of
the theorem we also need the compatibility conditions
(\ref{eq:fg},\ref{eq:ffg}). A more detailed account of this proof will be given
elsewhere.\\

At this point two remarks are in order.\\
First, It was noticed long ago \cite{MN2},
that any solution of the quantum Yang-Baxter equation (\ref{eq:yb}) leads
to solutions of the Tetrahedron equation (we consider here the case with
no dependence on shifts $(x)$) (\ref{eq:T}) as,
\begin{equation}
R_{11',22',33'}(u,v,w)\ =\ R_{12}(v,w)\ R_{1'3}(u,w)\ R_{2'3'}(u,v)
\label{eq:rt}
\end{equation}
However such solutions are degenerate in the sense that the partition
function of such a model will decompose into the product of three partition
functions of two-dimensional models associated to the planes $(u,v)$, $(u,w)$,
$(v,w)$. The solutions we propose here are not of this type since the existence
of a non-trivial $\Phi$ ensures precisely that the two-dimensional equations
such as (\ref{eq:FF2}) and hence (\ref{eq:yb}) are broken.\\
Second, the restricted Star-Triangle equations proposed in ref. \cite{BB1} are
likely to be very similar to our Pentagon equation. This point deserves further
study.\\

The next step is of course to find solutions $\Phi$ and $\Gamma$. In fact, if
we consider  eqs. (\ref{eq:sfi},\ref{eq:sg},\ref{eq:ufi},
\ref{eq:ug},\ref{eq:cg},\ref{eq:P},\ref{eq:fg},\ref{eq:ffg}) for functionals
having vector indices on links (and eventually on surfaces), solutions are
already at hand. Their are given by Conformal Field Theories or by
Topological Field Theories in the sense of Turaev and Viro \cite{TV}. In this
case $\Phi$ satisfies a usual Pentagon equation and $\Gamma$ is the
product of Kroenecker delta's. However in that case the $R$-matrix turns out
to be non-invertible. Moreover the model is topological \cite{MR1}. In fact
in that situation the proof of the Tetrahedron equation is a trivial
consequence of the Turaev-Viro theorem. The problem of finding more
general solutions (in particular non-topological one's, and depending
on spectral parameters) is now under study.\\
\\

{\bf 4. Conclusion}\\
\\

Using a geometrical interpretation of the $R$-matrix solving the
Tetrahedron equation
as a transport operator acting in a space of functionals of surfaces, we have
obtained a decomposition of such an $R$-matrix in terms of more fundamental
objects $(\Phi, \Gamma)$, $\Phi$ being the solution of the Pentagon equation
(\ref{eq:P}). This provides an explicit link between Pentagon and Tetrahedron
equations. We expect such a relation to be fruitful in the construction of
new solutions to the Tetrahedron equation. It also open the possibility of
extending the algebraic picture of Quantum Groups as given in \cite{Drin1} to
another algebraic structure suitable for Integrable Systems in three
dimensions. Finally, as can be expected from the fundamental relation
(\ref{eq:FFn}), it also relates three-dimensional Topological Field Theories
to a special case of the Tetrahedron equation.\\

\end{document}